\documentstyle[multicol,prb,aps,epsfig]{revtex}

\begin{document}

\twocolumn[\hsize\textwidth\columnwidth\hsize\csname @twocolumnfalse\endcsname

\title{Scaling of the conductance in a quantum wire}

\draft

\author{V.\ Meden$^1$, S.\ Andergassen$^2$,   
W.\ Metzner$^2$, U.\ Schollw\"ock$^2$, and K.\ Sch\"on\-hammer$^1$}
\address{$^1$Institut f\"ur Theoretische Physik, Universit\"at G\"ottingen, 
Tammannstr.\ 1, D-37077 G\"ottingen, Germany \\
$^2$Max-Planck-Institut f\"ur Festk\"orperforschung,
  Heisenbergstr.\ 1, D-70569 Stuttgart, 
Germany} 

\date{July 18, 2003}
\maketitle

\begin{abstract}
The conductance $G$ of an interacting nano-wire containing an 
impurity and coupled to non-interacting semi-infinite leads is studied 
using a functional renormalization group method. 
We obtain results for microscopic lattice models without any further 
idealizations. For an interaction which is turned on smoothly at the
contacts we 
show that one-parameter scaling of $G$ holds. If abrupt  
contacts are included we find power-law suppression of $G$ with 
an exponent which is twice as large as the one obtained for 
smooth contacts and no one-parameter scaling.  
Our results show excellent agreement with the analytically known 
scaling function at Luttinger liquid parameter $K=1/2$ and numerical 
density-matrix renormalization group data. 
\end{abstract}


\vspace{-0.6cm}

\vskip 2pc]
\vskip 0.1 truein
\narrowtext

The interplay of electron correlations and a
single impurity in one-dimensional electron systems
leads to striking
effects in the low-energy 
physics.\cite{LutherPeschel,Mattis,ApelRice,Giamarchi,KaneFisher} 
Using bosonization and
including a single impurity in an idealized manner the application of
a perturbative (in the strength of the impurity) renormalization 
group (RG) method to the resulting local sine-Gordon model (LSGM) 
led to a simple picture: at low energy scales 
physical observables behave as if 
the system is split in two chains with open boundary conditions at the 
end points.\cite{KaneFisher}
The bulk part of the model studied is known to capture the
universal bulk Luttinger liquid (LL) physics,\cite{Haldane} 
which is characterized by the interaction dependent LL parameter $K <
1$ (repulsive interaction). 
Considering the conductance $G$ as a function of the temperature $T$ and
the strength of the impurity $v$ it was argued that for fixed 
$K$ the RG flow from weak to strong impurity strength determines a  
scaling function $\tilde G_K(x)$ on which the data for different $T$ and 
$v$ can be collapsed (one-parameter scaling). Going
beyond the perturbative RG it was shown that scaling 
indeed holds for the LSGM.\cite{KaneFisher,Moon,Fendley} Applying the 
thermodynamic Bethe ansatz\cite{Fendley} (BA) $\tilde G_K$ 
was determined explicitly for $K=1/2$
and $K=1/3$.\cite{KaneFisher,Moon,Fendley}
In Refs.\ \onlinecite{KaneFisher,Moon,Fendley} the transport 
in fractional quantum Hall effect systems was studied and 
thus an infinite LL without leads was considered. Then in the
impurity free case the conductance is renormalized by the 
interaction, $G^\ast= K e^2/h$.
If the connection of a finite LL quantum wire to
non-interacting leads is modeled by a position dependent LL parameter
$K(x)$ which takes the value one in the leads (local LL description), 
the conductance
is given by $G^\ast=e^2/h$ in the impurity free case.\cite{Schulz,Maslov} 
The question how the one-parameter scaling of the conductance of 
the system with an impurity is modified in the
presence of leads was not addressed before.

Using numerical methods it was shown that the LSGM captures the behavior 
observed in a microscopic lattice model in the limits of weak and strong 
impurities.\cite{EggertAffleck} Based on these results it is generally
believed that (i) one-parameter scaling holds for a large class of models
of correlated electrons with a single impurity and (ii) 
the data always scale on the universal scaling function $\tilde G_K$ found 
within the LSGM.
The issue of universality is of special
importance since transport experiments have been interpreted in terms 
of scaling.\cite{experiments}  

Here we study the transport through an interacting nano-wire with an 
impurity connected to non-interacting semi-infinite leads. The wire is
modeled by the lattice model of spinless fermions with nearest-neighbor 
interaction and a hopping impurity. Using a fermionic functional 
RG method\cite{VM1} we are able to calculate $G$. 
We show that for a single impurity, if the interaction is 
turned on very smoothly starting at the contacts 
and no one-particle scattering terms (see Eq.\
(\ref{spinlessfermdef})) at the contacts are considered 
(smooth contacts) 
one-parameter scaling holds 
also in the presence of leads.
For $K=1/2$ the data perfectly match the LSGM curve 
$\tilde G_{K=1/2}$ provided the impurity free conductance 
$G^\ast$ 
is taken to be $e^2/h$ instead of $Ke^2/h$\cite{KaneFisher}
as appropriate for systems with leads.  
The latter gives us confidence that, despite the 
approximate treatment of the 
interaction 
our approach captures the relevant physics. 
This is further supported by a
comparison to numerical data for the conductance obtained
by the density-matrix renormalization group (DMRG) method.  
In the more generic case with abrupt contacts, 
i.e.\ if the interaction is turned on more rapidly 
or one-particle scattering terms at the contacts are included, 
we find a low-energy power-law suppression of $G$ with an exponent 
which is twice as large as the one obtained for 
smooth contacts 
and no one-parameter scaling. Evidently in most 
experiments with quantum wires the contacts are  abrupt. 

The model we study is given by the Hamiltonian
\begin{eqnarray}
H & = & - \sum_{j=-\infty}^{\infty} \hspace{-0.15cm}^{'}
%\sum \nolimits'
\left( c_j^{\dag} c_{j+1}^{} +
  \mbox{h.c.} \right)  \nonumber \\ && - t_l 
\left( c_{0}^{\dag} c_{1}^{} + \mbox{h.c.}
\right) - t_r \left( c_{N_W}^{\dag} c_{N_W+1}^{} + \mbox{h.c.}
\right) \nonumber \\ &&  -\rho  
\left( c_{N_W/2}^{\dag} c_{N_W/2+1}^{} + \mbox{h.c.}
\right) \nonumber \\ && + \sum_{j=1}^{N_W-1} U_j 
\left(n_j- \frac{1}{2} \right)  
\left(n_{j+1}- \frac{1}{2}\right) \; , 
\label{spinlessfermdef}
\end{eqnarray}
in standard second-quantized notation. The sum in the first line runs
over all lattice sites $j$ excluding $j=0$, $N_W/2$, and $N_W$, which
is indicated by the prime. 
The hopping matrix element and the lattice constant are set to 1. 
We consider an even number $N_W$ of lattice sites in the 
interacting region (the wire). A single hopping impurity of strength 
$\rho \leq 1 $ and modified hopping matrix elements $t_{l/r} \leq 1$ in and 
out of the wire are included.
Here we focus on the half-filled system. 
The
nearest-neighbor interaction $U_j$ close to the contacts at sites $1$
and $N_W$ is assumed to be spatially varying. The constant 
bulk value of the interaction is denoted by $U$. 
The bulk LL parameter is given by 
$K=\left[ \frac{2}{\pi} \arccos{ \left( - \frac{U}{2} 
\right)}\right]^{-1}
$, for $U\leq 2$,
as follows from the BA 
solution.\cite{HaldaneII,footnotesmallestK}

In linear response and at $T=0$ the
conductance can be determined from the one-particle Green
function of the interacting wire 
taken at the chemical potential and 
calculated in the presence of the non-interacting semi-infinite 
leads.\cite{Langer} 
For the half-filled case we obtain 
$G=4 e^2 t_l^2 t_r^2 \left| G_{N_W,1}(0) \right|^2/h$, with the Green function 
$G_{N_W,1}$.

In our earlier applications of the RG method\cite{VM1,VM2} 
we neglected the flow of the two-particle vertex 
and considered the flow of the self-energy only. Within this
approximation the LL exponents relevant for the impurity
problem turned out to be correct to leading order in the
interaction. We here go 
beyond this and include the flow of the vertex in an approximate 
way. 
We replace the three-particle 
vertex by its initial value $0$. The flow
equation for the two-particle vertex then reads
 \begin{eqnarray*}
&& \partial_{\Lambda} \Gamma^{\Lambda}(\alpha, \beta; \gamma, \delta) 
= \\ &&    - \mbox{ Tr}\, \Big\{ P^{\Lambda} 
\Gamma^{\Lambda}(\ldots ,  \ldots ;
\gamma, \delta ) \left[ G^{\Lambda}\right]^{t}   
\Gamma^{\Lambda}(\alpha, \beta ; \ldots
,\ldots ) \Big\}
\\ && - \mbox{ Tr}\, \Big\{ P^{\Lambda} 
 \Gamma^{\Lambda}(\alpha, \ldots ;
\gamma, \ldots) 
G^{\Lambda}   \Gamma^{\Lambda}(\beta, \ldots ;
\delta, \ldots)  \\ &&  - \left[ \alpha \leftrightarrow \beta \right] 
 - \left[ \gamma \leftrightarrow \delta \right] + 
 \left[ \alpha \leftrightarrow \beta ,  \gamma \leftrightarrow \delta 
\right]  \Big\}
 \; .
\end{eqnarray*}
The Greek letters stand for the quantum
numbers of the basis in which the problem is considered 
and the Matsubara frequencies.  
On the right hand side $\Gamma^{\Lambda}$ is understood as a matrix 
in the variables which are not written. 
$ P^{\Lambda} = G^{\Lambda} \left( \partial_{\Lambda} \left[ 
  G^{0,\Lambda}\right]^{-1} \right) G^{\Lambda}$,
with $G^{\Lambda} = \left(\left[ G^{0,\Lambda}\right]^{-1} -
  \Sigma^{\Lambda} \right)^{-1} $
and the cut-off dependent self-energy $\Sigma^{\Lambda}$.
$G^{0,\Lambda}$ is the non-interacting impurity free propagator
supplemented by an infrared cut-off.
As before\cite{VM1} we use a frequency cut-off $G^{0,\Lambda} =
\Theta(|\omega|- \Lambda) G^{0}$, with $\Lambda \in ]\infty,0]$. 
In the flow of the vertex we replace $G^{\Lambda}$ by $G^{0,\Lambda}$. 
We neglect the frequency dependence of the vertex which leads to
a frequency independent self-energy. 
Due to this the bulk LL properties of the model are only partially 
captured by our approximation. In particular we miss the bulk anomalous
dimension which is small compared to the impurity 
contribution included in our approach. 
We parameterize $\Gamma^{\Lambda}$ by an effective
nearest-neighbor interaction with a renormalized amplitude $U^{\Lambda}$,
whose flow is determined by projecting onto the Fermi 
points.
In this way the fixed point coupling is
guaranteed to be correct to order $U^2$. The coupling to the leads 
is neglected. 
Within these approximations the flow equation for $U^{\Lambda}$ 
closes.   
In the thermodynamic limit it can be integrated analytically leading to
\begin{eqnarray}
\frac{U^{\Lambda}}{U} = \left(1+\frac{\Lambda U}{2 \pi}- \frac{U}{2\pi}
  \frac{2+ \Lambda^2}{\sqrt{4+ \Lambda^2}} \right)^{-1} \; .
\label{ULambda}
\end{eqnarray}
Details of this approximation scheme  are presented elsewhere.\cite{Sabine} 
The one-particle Green function of the 
interacting wire in the presence of the semi-infinite leads and thus
the conductance can then be
determined numerically by integrating Eqs.\ (14) and (15) 
of Ref.\ \onlinecite{VM1}  
using $U^{\Lambda}$ instead of $U$. 
On the right hand side of Eq.\ (14) the self-energy on sites 
$1$ and $N_W$ has to be modified: due to the coupling to the leads
terms which can be expressed by $t_{l/r}$
and the $t_{l/r}=0$ Green function of the leads at sites $0$ and 
$N_W+1$ have to be added. The initial condition for 
$\Sigma^{\Lambda}$ is given by the impurity potential. 
Below we show that the above approximation scheme leads to 
exceptionally good results.

\begin{figure}[hbt]
\begin{center}
\vspace{0.5cm}
\leavevmode
\epsfxsize7.4cm
\epsffile{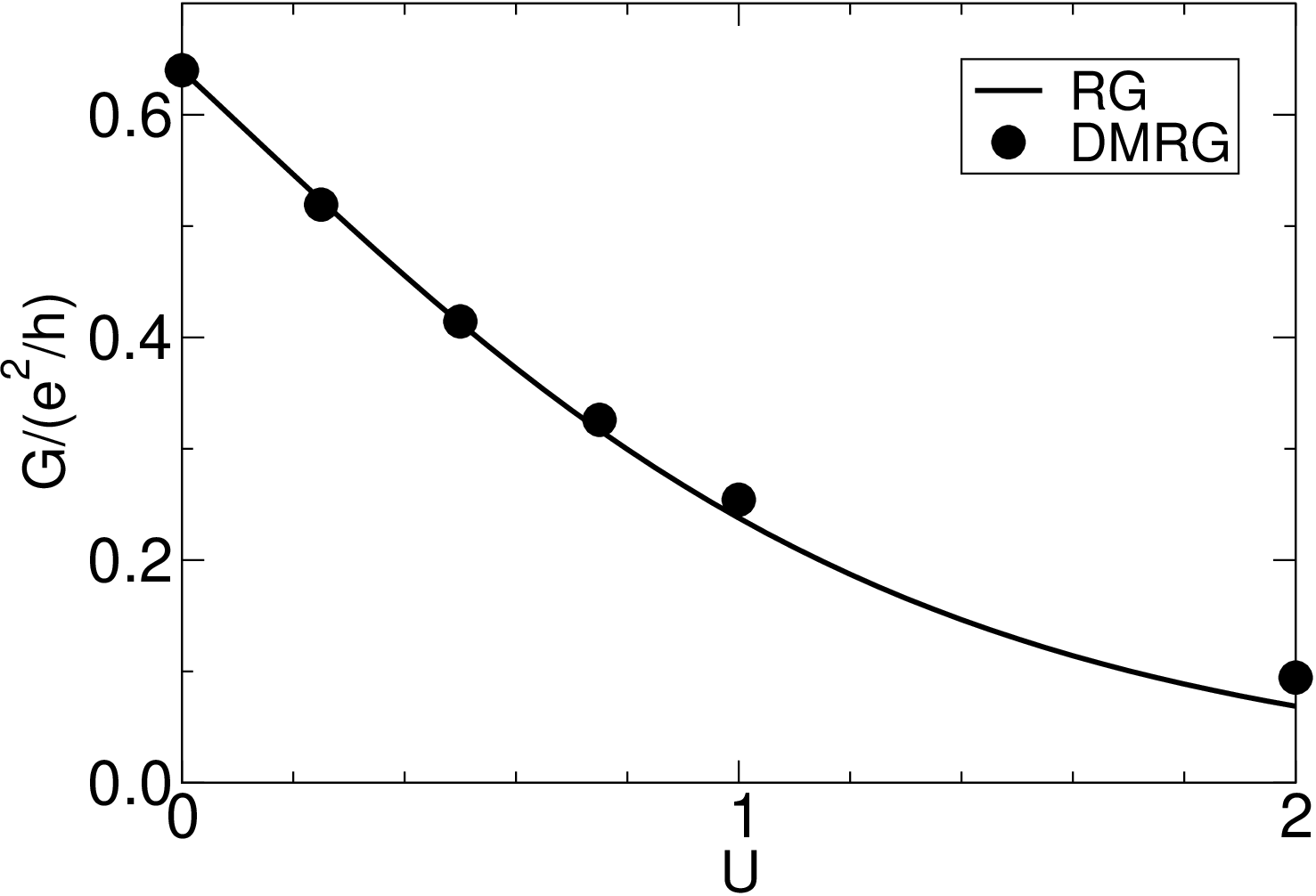}
\vspace{0.4cm}
\caption{Conductance as a function of $U$ for
  $N_W=12$, $\rho=0.5$, $t_{l/r}=1$. The interaction is turned on sharply.}
\label{fig1}
\vspace{-0.4cm}
\end{center}
\end{figure}

It was very recently 
suggested that for interacting systems the conductance through 
a wire can be determined from the persistent current which is observed 
in the presence of a magnetic flux piercing a ring in which the wire 
is embedded.\cite{Sushkov,Molina,Ramsak} 
In Ref.\ \onlinecite{VM2} two of us presented 
results which strongly support this prediction. This enables
us to obtain the conductance using the DMRG method 
(for details see \cite{VM2} and \cite{Molina}) 
and by comparison to verify the quality of our approximate treatment of the
interaction. 
In Fig.\ \ref{fig1} we present RG (solid line) and
DMRG\cite{DMRGsize} (circles) results for $G/(e^2/h)$ as
a function of $U$ for $U_j=U$, $j=1,2,\ldots,N_W-1$, $t_{l/r}=1$, 
and $\rho=0.5$.  
The good agreement of the data for interactions as large
as $U=2$ gives us confidence that the functional RG with the
approximations discussed above provides data for $G$ which are very
close to the exact value. To determine $U^{\Lambda}$ we numerically
integrated the finite size flow equation for the vertex, but even
for $N_W=12$ the difference between using this and the $N_W \to \infty
$ result of Eq.\ (\ref{ULambda}) is only marginal.

\begin{figure}[hbt]
\begin{center}
\vspace{0.5cm}
\leavevmode
\epsfxsize7.4cm
\epsffile{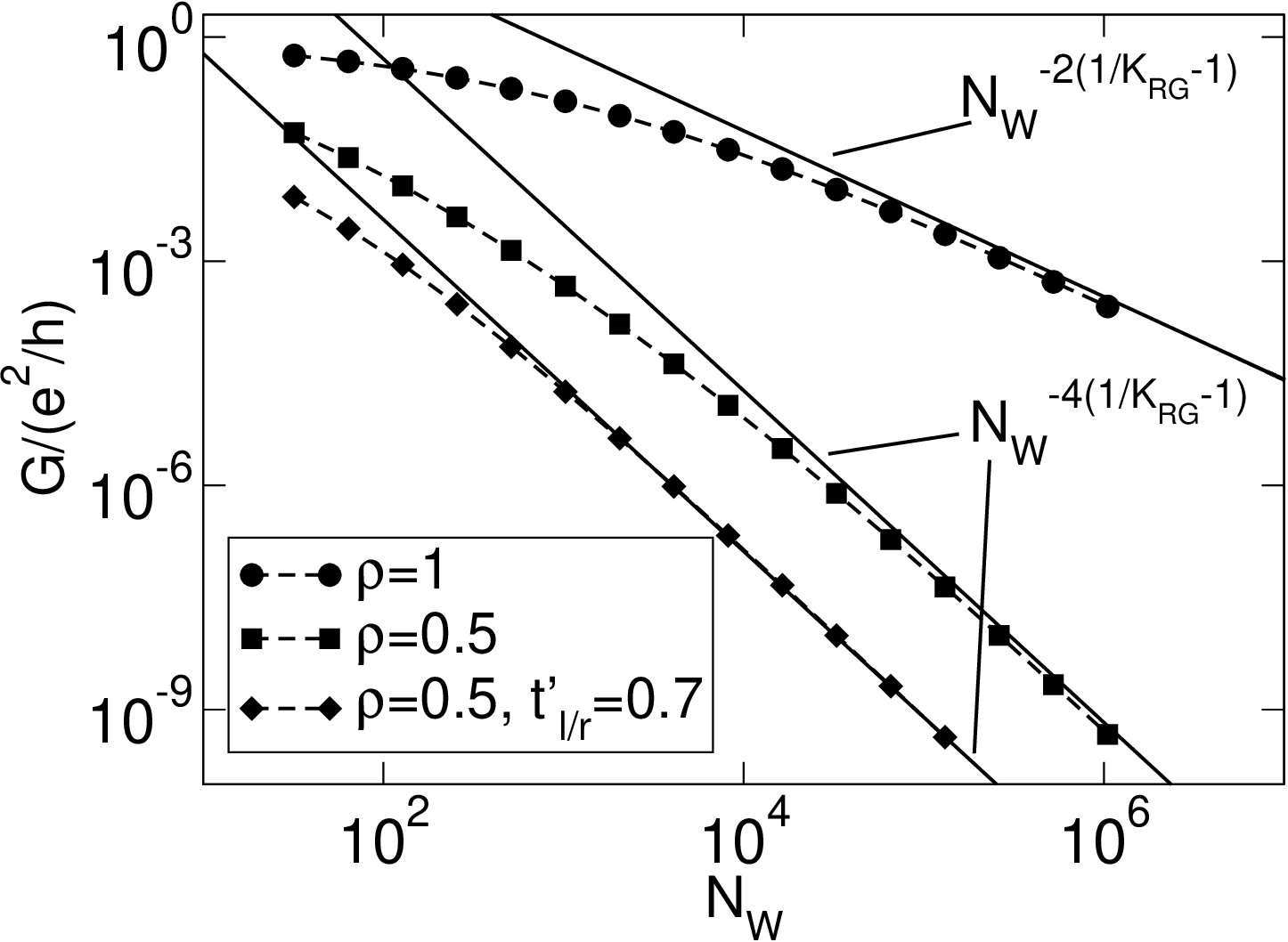}
\vspace{0.4cm}
\caption{Conductance as a function of $N_W$ for $U=1.5$, 
  $\rho=1$, $t_{l/r}=1$ (impurity free; circles), 
  $\rho=0.5$, $t_{l/r}=1$  (squares), and
  $\rho=0.5$, $t_{l/r}=0.7$ (diamonds). The
  interaction is turned on sharply.}
\label{fig1neu}
\end{center}
\end{figure}

Assuming an interaction which is turned on (and
off) sharply (in space), i.e.\ $U_j=U$ for $j=1,2,\ldots,N_W-1$, leads 
already in the absence of single particle scattering terms
(i.e. $t_{l/r}=\rho=1$) to a conductance which depends on $U$ and 
$N_W$.\cite{Oguri2,Molina,VM2} For large $N_W$, 
i.e.\ at low energy scales, $G$ goes to zero with a power-law 
as shown in Fig.\ \ref{fig1neu} (circles). This strong influence 
of an abrupt contact has not been
discussed before. To unambiguously determine the 
exponent exceptionally large wires of up to $10^6$ lattice 
sites have to be considered.\cite{fastprogram} 
For smaller $U$ even larger $N_W$ are required. We find that
$ G \sim N_W^{2(1-1/K)}$. 
Since our approach involves an approximate 
treatment of the interaction we only obtain an approximation 
$K_{\rm RG}$ for $K$. For $U=1.5$ we find 
$K_{\rm RG} = 0.643$, in excellent agreement 
with the BA result $K = 0.649$.  
The above exponent is exactly the one found
in the LSGM including one impurity\cite{KaneFisher}
if $T$ is replaced by $1/N_W$ (see below). The other curves of Fig.\
\ref{fig1neu} will be discussed later.
To avoid the suppression of $G$ due to the contacts and to 
investigate the role of a single impurity 
in the wire we now turn on the interaction smoothly: 
$U_j=U [\pi/2+ \arctan{(s [j-j_s/2])}]/\pi$ for 
$j=1,2,\ldots,j_s-1$, $U_j=U$ for $j=j_s,\ldots,N_W/2$, and
$U_{N_W-j}=U_j$ for $j=N_W/2+1,\ldots,N_W-1$.
The larger $N_W$ and $U$ the smoother $U_j$ has to be varied to
obtain the impurity free conductance for $t_{l/r}=\rho=1$.
This procedure enables us to quantify   
the terms ``perfectly'' and ``adiabatically'' connected 
used in the field theoretical modeling of transport through an impurity
free LL.\cite{Schulz,Maslov} 
In the absence of an impurity 
we find $G^\ast = e^2/h$ 
in agreement with the local LL description mentioned 
earlier.\cite{Schulz,Maslov}
For the results presented below we choose $s$ and 
$j_s$ such that for $t_{l/r}=\rho=1$ the  
relative deviation of the conductance from $e^2/h$ 
is less than $2 \times 10^{-4}$. For the system sizes 
considered in the following ($N_W \geq 256$) we do not expect 
the local change of the interaction strength over a few lattice 
sites to have a relevant effect on the flow of the bulk two-particle 
vertex and thus neglect this local change in the flow of the vertex. 
It turned out that as long as the switching on of the interaction is
smooth enough and the bulk part of the wire is large compared to the 
switching region, $G$ is independent of the details of the
switching procedure as expected. To determine $U^{\Lambda}$ we
use Eq.\ (\ref{ULambda}) obtained for $N_W \to \infty$. 

In the LSGM the conductance is studied
as a function of temperature.\cite{KaneFisher,Moon,Fendley} Here we limit 
ourselves to $T=0$ but treat wires of finite length. 
We expect 
that the temperature scaling can directly be translated into 
a scaling in $1/N_W$ and thus 
study $G$ as a function of $1/N_W$. For a certain value of $U$ we later 
confirm this by a direct comparison of the scaling function 
calculated by us with $\tilde G_{K=1/2}$ obtained within the LSGM. 
The variable in which scaling is expected is 
$N_W/N_0$, where $N_0$ denotes a 
non-universal length scale, i.e.\ $N_0$ depends 
on the details of the model and its parameters [here $N_0=N_0(U,\rho)$].

\begin{figure}
\begin{center}
\vspace{0.5cm}
\leavevmode
\epsfxsize7.4cm
\epsffile{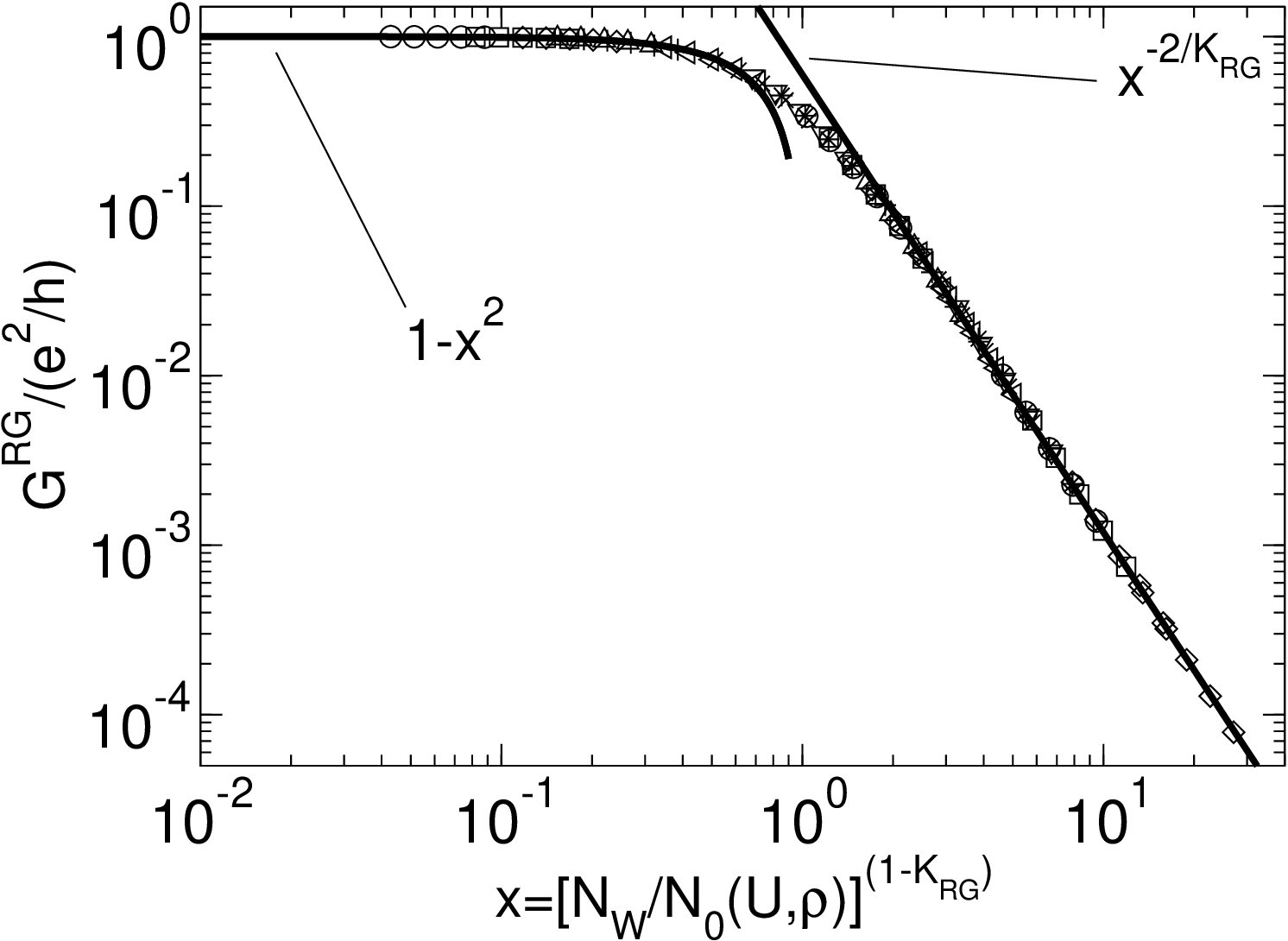}
\vspace{0.4cm}
\caption{Scaled conductance for an interaction which is turned on
  smoothly with the bulk strength $U=1$ and $t_{l/r}=1$. 
  Different symbols stand for
  different $\rho$ and $N_W=256,512,\ldots,4096$ in each case.}
\label{fig3}
\vspace{-0.4cm}
\end{center}
\end{figure}

For the universality of scaling to hold it is necessary that $G$ 
for small impurity strength, i.e.\ $1-\rho \ll 1$, and small 
$N_W$ scales as $1- G/(e^2/h) \sim N_W^{2(1-K)}$ and for large
impurity strength, i.e.\ $\rho \ll 1$, and large $N_W$ as $G/(e^2/h) 
\sim N_W^{2(1-1/K)}$.\cite{KaneFisher} 
Using the functional RG we find power-law 
behavior in both these limits with exponents which according to the
above two relations can be expressed consistently in terms of a single 
approximate LL parameter $K_{\rm RG}$. For example we find $K_{\rm
  RG}(U=0.5)=0.858$ and $K_{\rm RG} (U=1) =0.741$ both in excellent
agreement with the exact values $K=0.861$ and $K=0.75$ obtained from
the BA. 
To demonstrate one-parameter scaling we calculated $G$ for
$\rho=0.99,0.98,\ldots,0.95$, $\rho=0.9,0.85\ldots, 0.05$,
$N_W=256,512,\ldots,4096$, $j_s=22$, $s=2$, and various $U$. 
As an example Fig.\ \ref{fig3} shows that for $U=1$ the
data can be collapsed on a single curve $G^{\rm RG}/(e^2/h)$ with
$x=[N_W/N_0(U,\rho)]^{1-K_{\rm RG}}$ as a scaling variable. The
limiting behavior discussed above leads to the asymptotic $x$
dependence (solid lines) indicated in the figure. 

\begin{figure}[hbt]
\begin{center}
\vspace{0.5cm}
\leavevmode
\epsfxsize7.4cm
\epsffile{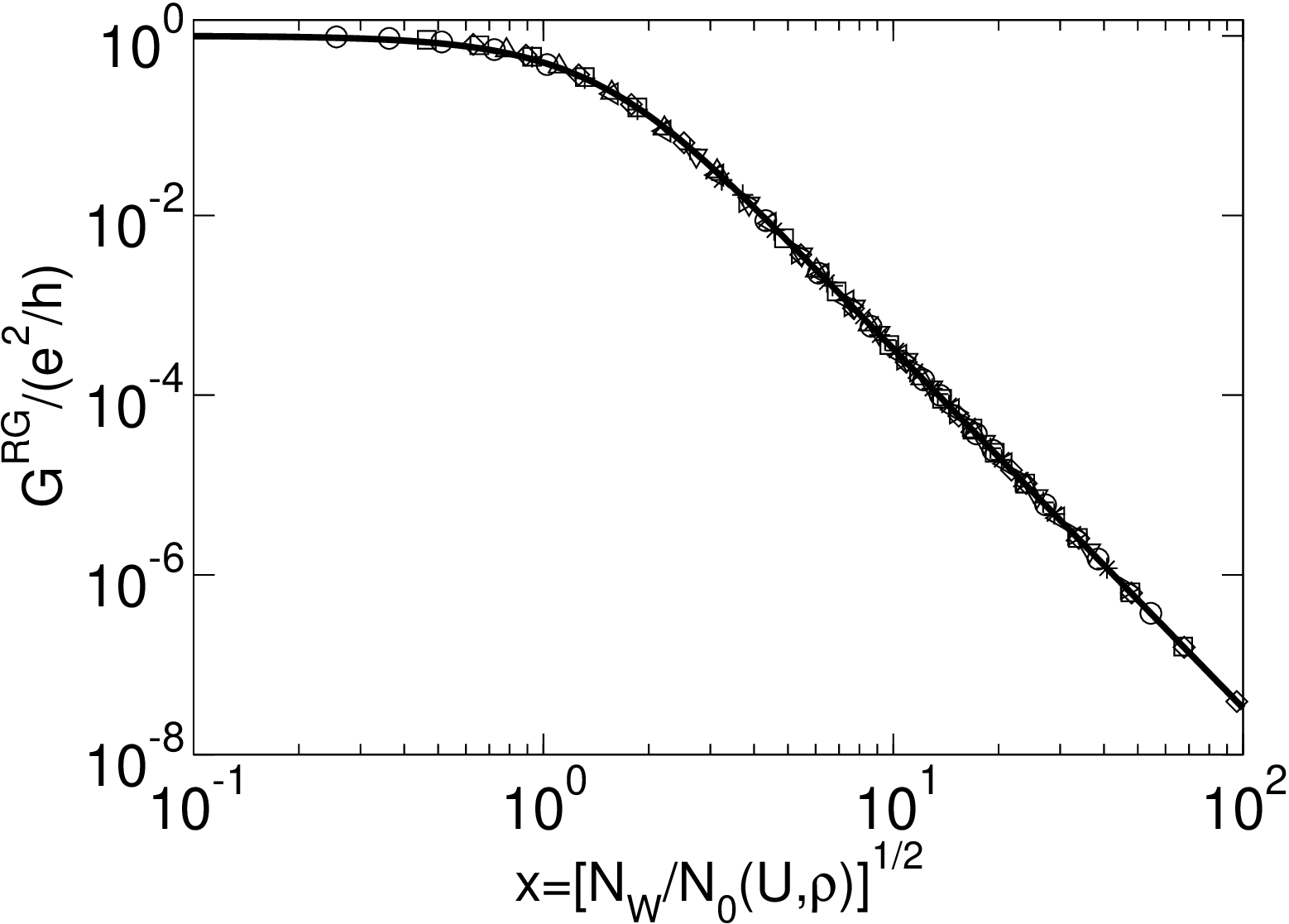}
\vspace{0.4cm}
\caption{The same as in Fig.\ \ref{fig3} but for $U=2.23$. The solid
  line is the $K=1/2$ scaling function of the LSGM.}
\label{fig4}
\end{center}
\end{figure}

After showing that we find  
one-parameter scaling for the lattice model
considered also in the presence of leads (provided the contacts are
assumed to be smooth)
we now compare the resulting scaling function 
to the one determined analytically within the LSGM for 
$K=1/2$.\cite{KaneFisher}
We find that $U=2.23$ leads to $K_{\rm RG} \approx 1/2$. This is
fairly close to $U=2$ which in the BA 
solution gives $K=1/2$. Within our approximation
the RG method does not capture the charge density wave ordering transition
at $U=2$ occurring in the exact treatment of the impurity free
model.\cite{footnotesmallestK}      
A comparison of the RG data ($\rho$, $N_W$, $j_s$, and $s$ as above)
with the LSGM
function (solid line) is presented in Fig.\ \ref{fig4}. 
The excellent agreement shows that in both physical situations 
the same scaling function is found
if the conductance
is divided by  $e^2/h$ instead of $Ke^2/h$, which is appropriate
when no leads are
present. The assumed equivalence of $T$ and $1/N$ scaling holds and 
our method captures the relevant physics quantitatively 
despite our approximate treatment of the interaction. 
Obviously (see Figs.\ \ref{fig3} and \ref{fig4}) the scaling function 
depends on the interaction, i.e.\ the LL parameter,
as expected from the LSGM. This has to be contrasted to the fermionic 
RG procedure of Ref.\ \onlinecite{MatveevGlazman}, where
the scaling function for arbitrary interaction turns 
out to be the non-interacting one, if the scaling variable $x$ is 
defined as above.

\begin{figure}
\begin{center}
\vspace{1cm}
\leavevmode
\epsfxsize7.4cm
\epsffile{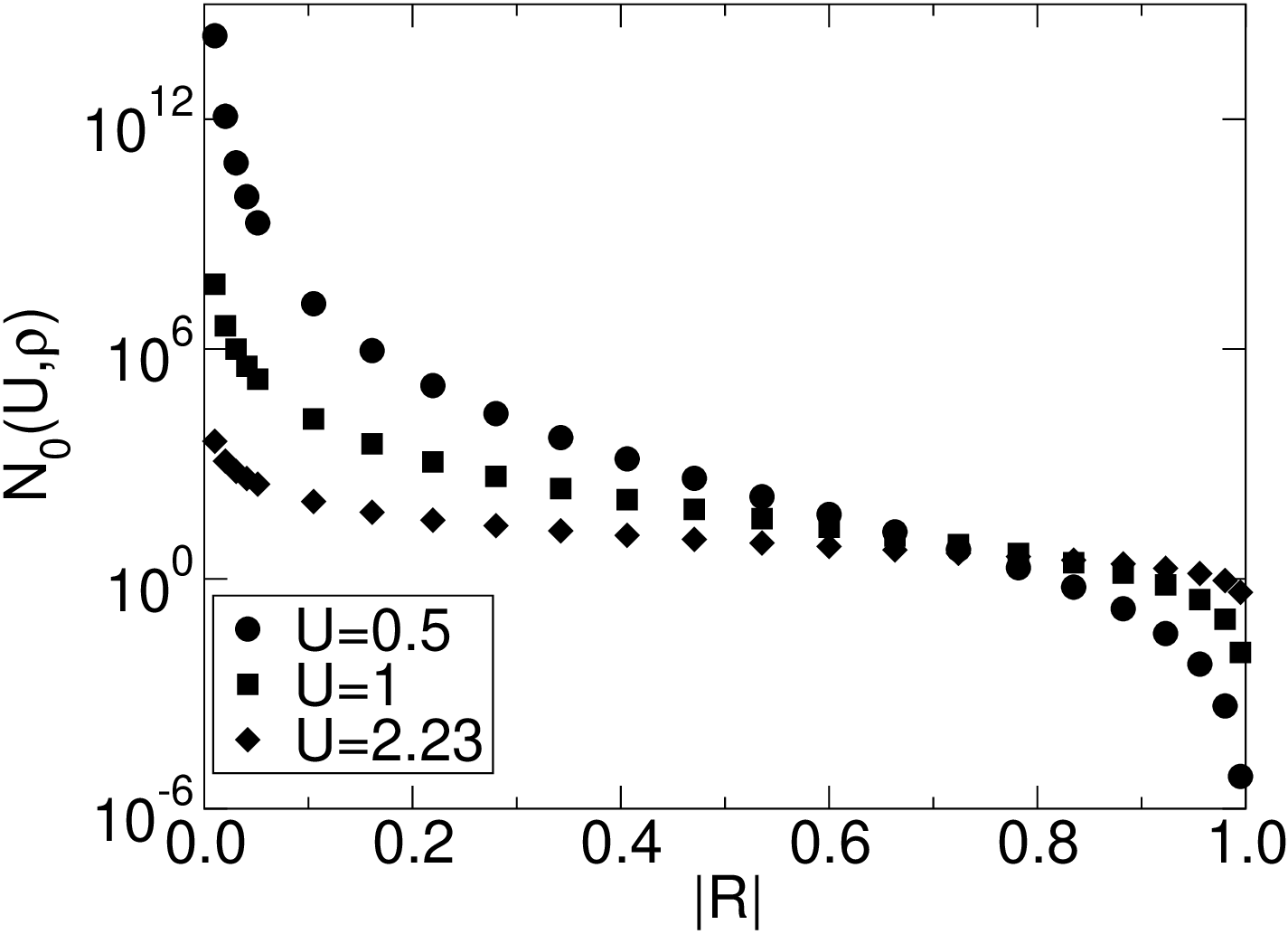}
\vspace{0.4cm}
\caption{$N_0$ as a function of $|R|$ for different $U$.}
\label{fig5}
\vspace{-0.4cm}
\end{center}
\end{figure}

For $U=0.5$, $1$, and $2.23$ the length scale $N_0$ is shown in Fig.\
\ref{fig5} as a function of the non-interacting reflection amplitude
$|R|=(1-\rho^2)/(1+\rho^2)$. For $N_0 \gg 1$ this scale provides a
measure for how large $N_W$ has to be for a given $\rho$ and $U$ 
before the  strong impurity limit is reached. 
Using a site impurity, a combination of site 
and hopping impurities, and different positions of the local impurity  
we have verified that scaling holds for 
generic types of impurities and that independently of the type of impurity 
for fixed $U$ the same scaling function is found. 

We next study the more generic case of a single impurity 
in an interacting wire taking abrupt contacts into account. 
They are modeled considering either $t_{l/r}<1$
or an interaction which is turned on sharply, 
i.e.\ $U_j=U$ for $j=1,2,\ldots,N_W-1$. For the latter
case $G$ as a function of $N_W$ is shown in
Fig.\ \ref{fig1neu} (squares) for $\rho=0.5$, $t_{l/r}=1$, 
and $U=1.5$. We find that 
the suppression of $G$ in the low energy limit again follows a
power-law but with the exponent $4(1-1/K)$ which is twice as large as
the one obtained for smooth contacts. The same holds if the
contacts are modeled by additional reduced $t_{l/r}$ 
independent of whether the interaction is turned on smoothly or 
sharply. In this case the asymptotic regime is reached 
for much smaller systems as can be seen in 
Fig.\ \ref{fig1neu} for $\rho=0.5$ and 
$t_{l/r}=0.7$ (diamonds). 
In both the above cases the data for different $\rho$ and $N_W$ 
cannot be collapsed on a single curve by a one-parameter scaling ansatz.  
Considering other types of impurities, asymmetric coupling to the
leads, and different positions of the local impurity we have verified 
that the exponent $4(1-1/K)$ is found generically.
We thus believe that the low-energy and resonant
tunneling\cite{KaneFisher} properties in most transport 
experiments on interacting quantum wires including a single 
impurity and coupled to non-interacting leads via contacts
are governed by the exponent $4(1-1/K)$ instead of $2(1-1/K)$.
One-parameter scaling cannot be expected to hold. 
This observation might be relevant for the attempts to interpret 
recent resonant tunneling experiments on carbon nanotubes.\cite{Postma} 

In summary, using a functional RG method we have shown that
in a lattice model of interacting one-dimensional 
electrons with a single impurity one-parameter scaling 
of the conductance holds even
in the presence of non-interacting leads if smooth contacts are 
considered, that is if the interaction is turned on smoothly and no
additional one-particle scattering terms at the contacts are included.
By comparison to numerical DMRG data and the $K=1/2$
scaling function of the LSGM we have verified that the approximate 
treatment of the interaction is valid up to fairly large interactions.
We have shown that in the impurity free case a sharp onset of the
interaction  leads to a power-law suppression of $G$ with an exponent
$2(1-1/K)$ which is the same as the one found for a single 
impurity within the LSGM. In the generic case of abrupt 
contacts and a single impurity in the wire we do not find 
one-parameter scaling and the low-energy properties are governed
by the exponent $4(1-1/K)$. The functional RG approach has 
the important advantage that it is very flexible and allows to 
determine the conductance 
for microscopic lattice models of correlated electrons without any 
further idealizations. The spin degree of freedom can  be 
included.\cite{Sabine} For further studies the combination of the 
fermionic lattice description and the functional RG approach 
allows for a more realistic microscopic modeling of contacts and leads.

\acknowledgments
We thank A. Rosch for urging us to clarify whether the
expected scaling behavior of the conductance is captured by
our RG approach.
U.S.\ is grateful to the Deutsche Forschungsgemeinschaft 
and Die Junge Akademie and V.M.\ to the Bundesministerium 
f\"ur Bildung und Forschung for support.


\begin{thebibliography}{0}
\bibitem{footnotesmallestK} This implies that the smallest $K$
  obtainable within the present model and for half-filling is $K=1/2$.
\bibitem{DMRGsize} To obtain the conductance using DMRG
  non-interacting leads have to be added and an extrapolation in 
  the length of the leads is required. Therefore the size of the
  interacting wire is fairly limited. \cite{VM2,Molina}
\bibitem{fastprogram} Due to an optimization of our code 
  suggested to us by T.\ Enss we are now able to treat much larger 
  chains compared to our earlier studies\cite{VM1}. E.g.\ obtaining $G$ for
  $N_{W} = 2^{20}$ takes roughly 15 minutes on a 2.4 GHz Pentium 4.
  We also thank X.\ Barnab\'e-Th\'eriault for discussions.
\end{thebibliography}
\end{document}